\documentclass[12pt,manuscript=article]{achemso}
\usepackage{graphics,epsfig,color,amsmath}
\usepackage{subcaption}
\usepackage{comment}
\title{Alchemical Transfer Approach to Absolute Binding Free Energy Estimation}

\author{Joe Z. Wu}
\affiliation{Department of Chemistry, Brooklyn College of the City University of New York, New York, NY}
\altaffiliation{These authors contributed equally to this work}
\alsoaffiliation{Ph.D. Program in Chemistry, The Graduate Center of the City University of New York, New York, NY}
\author{Solmaz Azimi}
\affiliation{Department of Chemistry, Brooklyn College of the City University of New York, New York, NY}
\altaffiliation{These authors contributed equally to this work}
\alsoaffiliation{Ph.D. Program in Biochemistry, The Graduate Center of the City University of New York, New York, NY}
\author{Sheenam Khuttan}
\altaffiliation{These authors contributed equally to this work}
\affiliation{Department of Chemistry, Brooklyn College of the City University of New York, New York, NY}
\alsoaffiliation{Ph.D. Program in Biochemistry, The Graduate Center of the City University of New York, New York, NY}
\author{Nanjie Deng}
\affiliation{Department of Chemistry and Physical Sciences, Pace University, New York, New York, 10038}
\author{Emilio Gallicchio}
\email{egallicchio@brooklyn.cuny.edu}
\affiliation{Department of Chemistry, Brooklyn College of the City University of New York, New York, NY}
\alsoaffiliation{Ph.D. Program in Chemistry, The Graduate Center of the City University of New York, New York, NY}
\alsoaffiliation{Ph.D. Program in Biochemistry, The Graduate Center of the City University of New York, New York, NY}

\begin{document}

\maketitle

\begin{abstract}
The Alchemical Transfer Method (ATM) for the calculation of standard binding free energies of non-covalent molecular complexes is presented. The method is based on a coordinate displacement perturbation of the ligand between the receptor binding site and the explicit solvent bulk,  and a thermodynamic cycle connected by a symmetric intermediate in which the ligand interacts with the receptor and solvent environments with equal strength. While the approach is alchemical, the implementation of ATM is as straightforward as for physical pathway methods of binding. The method is applicable in principle with any force field, it does not require splitting the alchemical transformations into electrostatic and non-electrostatic steps, and it does not require soft-core pair potentials. We have implemented ATM as a freely available and open-source plugin of the OpenMM molecular dynamics library. The method and its implementation are validated on the SAMPL6 SAMPLing host-guest benchmark set. The work paves the way to streamlined alchemical relative and absolute binding free energy implementations on many molecular simulation packages and with arbitrary energy functions including polarizable, quantum-mechanical, and artificial neural network potentials.
\end{abstract}

\section{\label{sec:intro}Introduction}

The binding free energy of a molecular complex is a rigorous thermodynamic measure of the degree of affinity of two molecules for each other. Measurements of binding free energies are useful in a wide range of chemical and medicinal applications ranging from drug discovery to chemical detection and toxicology. The ability to estimate binding free energies by computational modeling adds an important dimension to this probe.\cite{simonson2016physical} Relative binding free energy models are widely used, for example, in drug lead optimization.\cite{Jorgensen2009} 

This work is concerned in particular with atomistic models of the absolute binding free energy. These methods can be divided into two classes that are based on the nature of the thermodynamic path used to connect the bound and unbound states of the molecular complex.\cite{Gallicchio2011adv} Physical pathway methods define a spatial coordinate along which the reversible work for bringing the two molecules together is calculated. Conversely, alchemical methods connect the bound and unbound states by a series of unphysical intermediate states. The Single-Decoupling Method (SDM),\cite{Gallicchio2010,pal2019dopamine} for example, is based on progressively turning on the effective interaction between the ligand and the receptor with an implicit representation of the solvent. The related Double-Decoupling Method (DDM),\cite{Gilson:Given:Bush:McCammon:97} which is widely used to estimate the absolute binding free energy with an explicit representation of the solvent, estimates the binding free energy as the difference between the free energies of coupling the ligand to a pure solvent system and to the solvated receptor. 

Physical and alchemical binding free energy methods are characterized by distinct challenges and limitations. While it is broadly employed, the alchemical DDM approach is known to suffer from poor convergence and strong bias, especially for large and charged ligands, and these consequences are characterized by large and compensating decoupling free energies that result in the amplification of statistical and systematic errors in the binding free energy estimate.\cite{deng2018comparing} Another significant shortcoming of DDM is the large perturbation of the environment of the ligand in going from the solvated states to the vacuum intermediate state. Unless the ligand is conformationally restrained, the transition from a solvated state to vacuum can trigger substantial intramolecular conformational changes that relax slowly to the bound or unbound configurations. Other technical limitations of DDM stem from differences in the composition and size of the molecular systems used for the two decoupling legs,\cite{rocklin2013calculating} and from the inconsistent treatment of long-range interactions.\cite{pan2017quantitative,oostenbrink2020charge} Even for relatively simple systems, these and other alchemical transformations have to be conducted with care to avoid singularities and slow convergence. It is recommended for example, to couple electrostatic interactions separately from other interactions,\cite{lee2020improved} and to employ customized soft-core pair potentials\cite{Steinbrecher2007} to avoid end-point singularities. 

Because they are often based on modifying the parameters and the form of the energy function,  software implementations of alchemical binding free energy methods also tend to be significantly complex and require in-depth knowledge of the data structures of the target molecular simulation package. For example, core energy routines are usually customized to implement the specific modified pair potentials that represent the interaction of the ligand with the rest of the system. These modified pair potentials depend on the alchemical progress parameter (generally denoted by $\lambda$) which, together with other alchemical variables, become additional system parameters. Methods such as Thermodynamic Integration (TI)\cite{Mey2020Best,lee2020alchemical} require additional routines to implement the calculations of the gradients of the energy function with respect to the parameters that are alchemically transformed. The implementation of alchemical transformations involving many-body potential terms, where the $\lambda$ dependence affects more than individual pair interaction energies, is particularly challenging. These include alchemical applications with polarizable potentials,\cite{harger2017tinker} Ewald long-range electrostatic treatments,\cite{Darden:93} implicit solvent models,\cite{Gallicchio2009} as well as some conventional intramolecular potential terms.\cite{zou2019blinded} 

Physical pathway methods\cite{Woo2005,limongelli2013funnel,deng2018comparing} address some the limitations of DDM by physically moving the ligand from the solvent bulk to the binding site. Physical pathway binding free energy calculations are seen as preferable over DDM for large and charged ligands because they are typically performed as one continuous path in a single solvent box without transferring the ligand to a different phase. In addition, software implementations of physical pathway methods do not require as much customization of the underlying molecular simulation package as alchemical methods do. The primary limitation of physical pathway methods is the high computational cost due to the requirement of equilibrating the complex at many intermediate receptor-ligand separations that might not be of interest. Because they require a physical exit and entry channel, physical pathway methods are also not generally applicable to occluded binding sites.\cite{cruz2020combining}


Building upon on our Single Decoupling Method (SDM) for absolute binding free energy estimation\cite{kilburg2018assessment,pal2019perturbation} implemented in OpenMM,\cite{eastman2017openmm} we have been investigating ways to streamline alchemical calculations with explicit solvation. SDM, which has been designed for binding free energy calculations with implicit solvation, computes the alchemical perturbation energy by translating the ligand from the solvent medium to the receptor binding site, rather than attempting to selectively turn-on and turn-off individual ligand-receptor interactions. The approach treats all interactions in one concerted step and employs the standard molecular mechanics force field without soft-core pair potentials. End-point singularities are addressed by a suitable non-linear alchemical energy function.\cite{pal2019perturbation} We have recently shown that the same approach is applicable to the estimation of the concerted hydration free energies of drug-sized solutes in water droplets with explicit solvation.\cite{khuttan2021single} 

In this work, we extend this concerted alchemical scheme to the calculations of absolute binding free energies with explicit solvation. In the resulting alchemical scheme, called the Alchemical Transfer Method (ATM) (Figure \ref{fig:atm_vs_ddm}B), the unbound and bound states of the molecular complex are related by a translation vector that brings the ligand from the solvent bulk to the receptor binding site in a single solvent box. We show that the proposed method addresses some of the aforementioned challenges of binding free energy calculations by exploiting the best characteristics of the alchemical and physical methods. Like DDM, the method is based on alchemical transformations that aim conformational sampling only in the solvent bulk and the receptor binding site, and like tphysical methods, ATM is based on moving the ligand in physical space, in a single simulation box, and without transferring the ligand to vacuum. The ATM method, implemented as a freely available plugin of the OpenMM molecular simulation package, did not require any modifications of the OpenMM core energy routines.  We validate the ATM approach on a rigorous benchmarking dataset developed by Rizzi et al.\cite{rizzi2020sampl6}

\section{\label{sec:methods}Theory and Methods}

\subsection{Alchemical Transformations}

Free energy changes are estimated using alchemical transformations based on a $\lambda$-dependent potential energy functions $U_\lambda (x)$ that brings the system from an initial state at $\lambda=0$, described by the potential function $U_0(x)$, to a final state at $\lambda=1$, corresponding to the  potential function $U_1(x)$. 

For each transformation, the alchemical potential energy is expressed as
\begin{equation}
U_{\lambda}(x)=U_{0}(x)+W_{\lambda}(u)\label{eq:pert_pot}
\end{equation}
where $x$ represents the set of atomic coordinates of the system,
\begin{equation}
u(x)= U_{1}(x)-U_{0}(x) \label{eq:sc-binding-energy}
\end{equation}
is the perturbation energy, and $W_{\lambda}(u)$ is the generalized softplus alchemical perturbation function
\begin{equation}
  W_{\lambda}(u)=\frac{\lambda_{2}-\lambda_{1}}{\alpha}\ln\left[1+e^{-\alpha(u_{\rm sc}(u)-u_{0})}\right]+\lambda_{2}u_{\rm sc}(u)+w_{0} .
  \label{eq:ilog-function}
\end{equation}
The parameters $\lambda_{2}$, $\lambda_{1}$, $\alpha$, $u_{0}$, and $w_{0}$ are functions of $\lambda$ (see Computational Details), and 
\begin{equation}
  u_{\rm sc}(u)=
\begin{cases}
u & u \le u_0 \\
(u_{\rm max} - u_0 ) f_{\rm sc}\left[\frac{u-u_0}{u_{\rm max}-u_0}\right] + u_0 & u > u_0
\end{cases}
\label{eq:soft-core-general}
\end{equation}
with
\begin{equation}
f_\text{sc}(y) = \frac{z(y)^{a}-1}{z(y)^{a}+1} \label{eq:rat-sc} \, ,
\end{equation}
and
\begin{equation}
    z(y)=1+2 y/a + 2 (y/a)^2
\end{equation}
is the soft-core perturbation energy. The soft-core function is monotonic map that avoids singularities near the initial state of the alchemical transformation at $\lambda=0$ without affecting the distribution of perturbation energies at the final state at $\lambda=1$. As expressed in Eq.\ (\ref{eq:soft-core-general}), the soft-core perturbation energy is designed to cap the perturbation energy $u(x)$ to a maximum positive value $u_{\rm max}$ and to be equal to the perturbation energy when this is below a cutoff value $u_0$. The $u_0$ cutoff is selected to be sufficiently large so that $u_{\rm sc}(u) = u$ for all observed samples collected at the end state. The specific values of $u_0$, $u_{\rm max}$ and of the scaling parameter $a$ used in this work are listed in Computational Details.

In order to reproduce the desired end points, it is necessary that the alchemical perturbation function is defined such that $W_{0}(u)=0$ and $W_{1}(u)=u$ at $\lambda=0$ and $\lambda=1$, respectively. This requirement is satisfied by the linear function, $W_{\lambda}(u)=\lambda u_{\rm sc}(u)$, which is special case of the softplus function in Eq.\ (\ref{eq:ilog-function})
for which $\lambda_1 = \lambda_2 = \lambda$. The linear function is the standard choice for the alchemical perturbation function. As it can be verified from Eq.\ (\ref{eq:ilog-function}), in general the end-point requirement is satisfied whenever $\lambda_1 = \lambda_2 = 0$ at $\lambda=0$ and $\lambda_1 = \lambda_2 = 1$ at $\lambda = 1$.

\subsection{The Alchemical Transfer Method for Binding Free Energy Estimation}

\begin{figure}
  \begin{center}
(A) \begin{subfigure}{.4\textwidth}
\includegraphics[scale = 0.35]{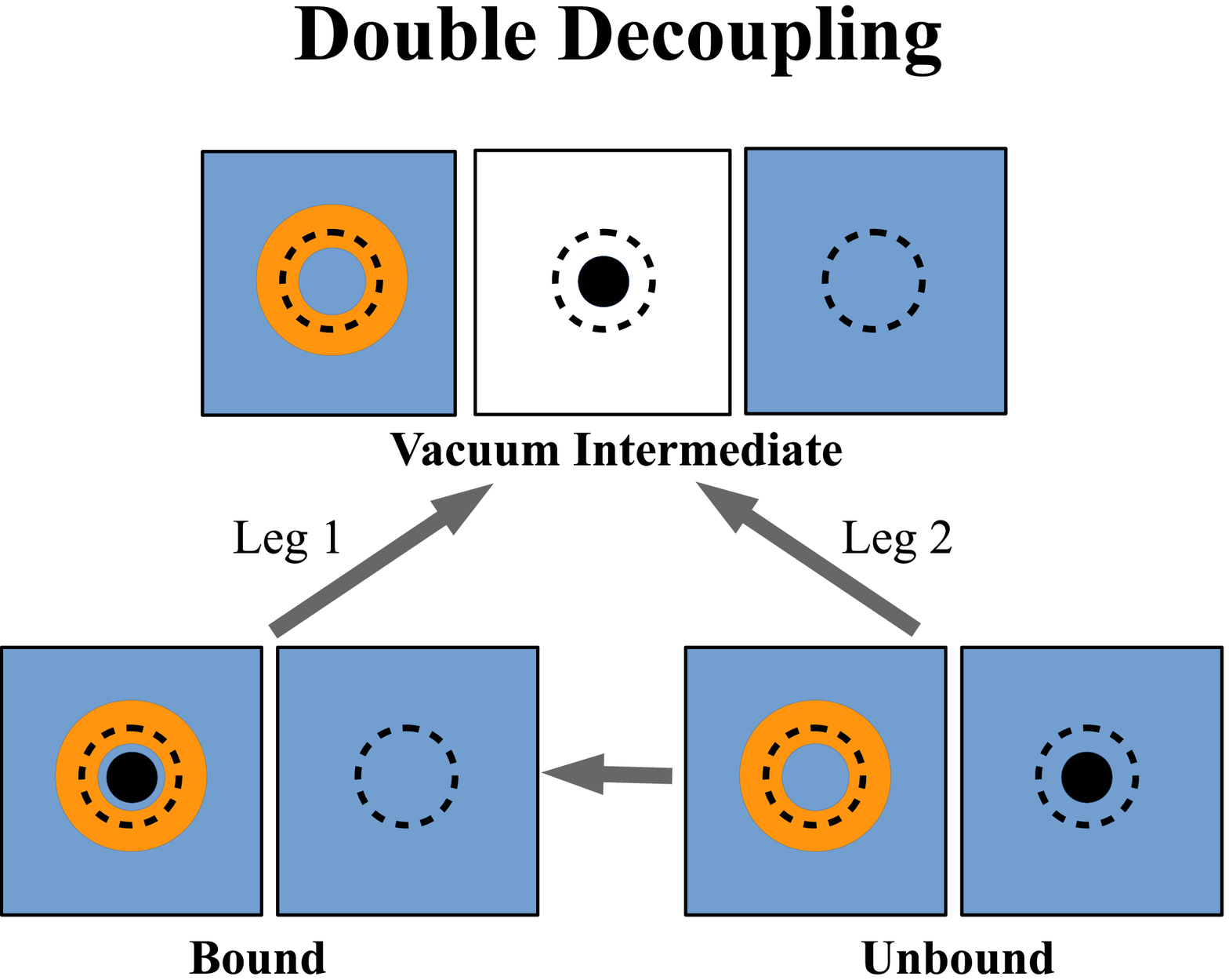}
\end{subfigure}
\hspace{0.5in}
(B) \begin{subfigure}{.4\textwidth}
\includegraphics[scale = 0.35]{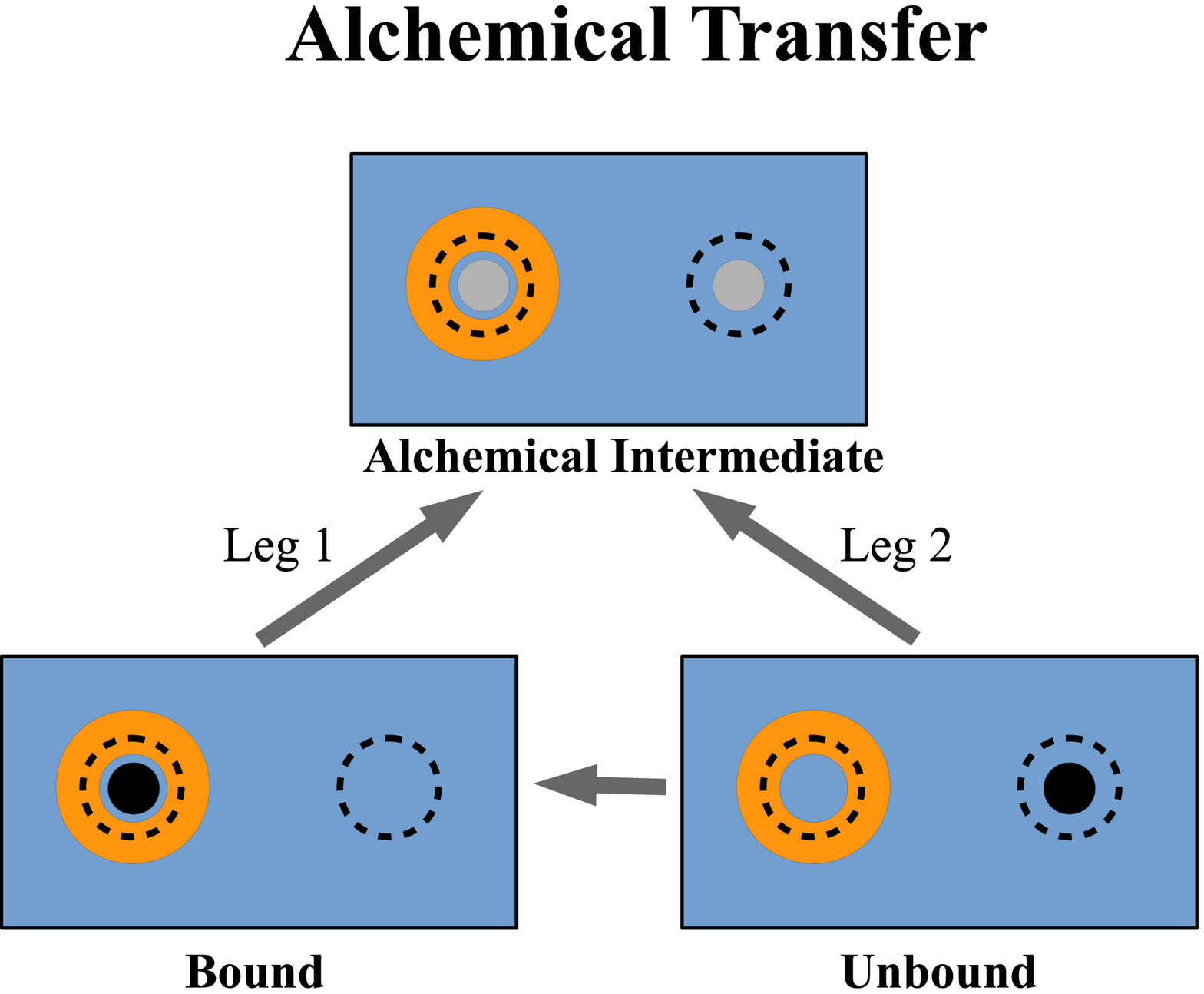}
\end{subfigure}
\caption{\label{fig:atm_vs_ddm}Schematic illustrations of (A) the double decoupling (DDM) and (B) the Alchemical Transfer (ATM) methods for the calculation of the binding free energy between a molecular receptor (orange doughnut) and a ligand (black circle). The dashed circle within the receptor represents the binding site region. The blue boxes represent the solvent. The unbound and bound end states for the two methods are considered thermodynamically equivalent. In both methods, the end states are transformed to a common intermediate state and the excess binding free energy is the difference of the free energy changes of the two legs, $\Delta G^\ast_b = \Delta G_2 - \Delta G_1$. Double decoupling defines an intermediate state in which the ligand is in vacuum (white). The alchemical transformations in the Alchemical Transfer Method are instead performed in the same solvent box and the ligand does not leave the solvated system.}
\end{center}
\end{figure}

Consider the non-covalent binding process between a receptor R  and a ligand L. The standard free energy of binding, $\Delta G^\circ_b$, is defined as the difference in free energy between the bound complex and the unbound components at the standard concentration of $C^\circ = 1$ M,
\begin{equation}
  \Delta G^\circ_b = \Delta G^\circ_{\rm site} + \Delta G^\ast_b .
  \label{eq:dgbind}
\end{equation}
where $\Delta G^\ast_b$ is the excess component, defined as the reversible work for transferring the ligand to the binding site region of the receptor of volume $V_{\rm site}$ from a region of the same volume in the solvent bulk (Figure \ref{fig:atm_vs_ddm}) plus a concentration-dependent term
\begin{equation}
  \Delta G^\circ_{\rm site} = -k_B T \ln  C^\circ V_{\rm site}
  \label{eq:dgsite}
\end{equation}
that corresponds to the free energy of transfer of a ligand molecule from an ideal solution at concentration $C^\circ$ to a region of volume $V_{\rm site}$ in the solvent.\cite{Gallicchio2011adv} In the remainder, we will focus on the calculation of the excess free energy component $\Delta G^\ast_b$ by alchemical molecular simulations.

The Double Decoupling Method (DDM)\cite{Gilson:Given:Bush:McCammon:97} has emerged as one of the gold standards for the alchemical calculation of absolute binding free energies in condensed phases.  DDM is based on the thermodynamic cycle illustrated in Fig.\ \ref{fig:atm_vs_ddm}A, whereby the bound and unbound states of the molecular complex are thermodynamically linked by an intermediate state in which the ligand is placed in vacuum. The excess binding free energy is expressed as the difference of the free energies of alchemically decoupling the ligand from the unbound (leg 1) and bound (leg 2) states into the intermediate vacuum state, as
\begin{equation}
  \Delta G^\ast_b = \Delta G_2 - \Delta G_1 \, .
  \label{eq:cycle_dg}
\end{equation}

\begin{figure}
  \begin{center}
    \includegraphics[scale = 0.35]{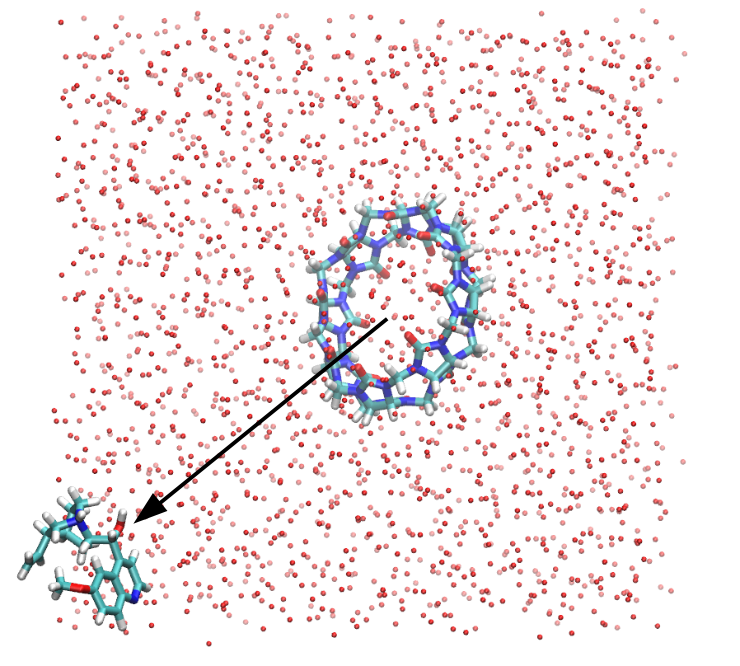}
    \end{center}
    \caption{\label{fig:cb8-g3}Illustration of the unbound state  of the complex between the CB8 host (center) and the G3 guest (lower left). The small red spheres represent the oxygen atoms of the water molecules. The displacement vector that translates the ligand from the binding site to the bulk solvent position is indicated.}
\end{figure}

The Alchemical Transfer Method (ATM) proposed here avoids the vacuum intermediate and requires only one molecular system.
As illustrated in Fig.\ \ref{fig:atm_vs_ddm}B, ATM is based on an alchemical intermediate in which the ligand interacts simultaneously with the solvent bulk and the receptor. We assume, without loss of generality, that there is a suitable coordinate frame attached to the receptor and that the binding site region (represented by the dashed circle in Fig.\ \ref{fig:atm_vs_ddm}) is fixed relative to this coordinate frame. Under these assumptions, every point in the binding site region maps to a unique point into a region of the same shape in the solvent bulk by means of a constant translation vector $h$, also at rest relative to the receptor coordinate frame (Figure \ref{fig:cb8-g3}). The bound state of the system is defined as any configuration in which the ligand is in the binding site region. Conversely, the unbound state of the system is defined as any configuration of the system in which the ligand is placed into the bulk solvent region (Fig.\ \ref{fig:atm_vs_ddm}B).

Under these assumptions, any configuration of the bound state maps to an unique configuration of the unbound state by a rigid translation of the ligand atoms by the vector $h$ (Figure \ref{fig:cb8-g3}). The reverse is also true. Any configuration of the unbound system maps uniquely to a configuration of the bound state by translation of the vector $-h$. Hence, the translation vector $h$ can be used as a perturbation parameter to connect, in a statistical thermodynamic sense, the bound and unbound states.

For example, consider leg 1 of the ATM cycle in Fig.\ \ref{fig:atm_vs_ddm}B. Denoting the potential energy function of the system as $U(x)$, with $x = (x_S, x_L)$ being the coordinates of the bound system, where $x_L$ are the coordinates of the ligand in the receptor binding site and $x_S$ are the coordinates of the receptor and the solvent (the surroundings), the potential energy functions $U_0$ and $U_1$ of the initial and final states of the leg (the bound and unbound states in this case) are respectively
\begin{equation}
U_0(x_S, x_L) = U(x_S, x_L)
\label{eq:pot_bound_1}
\end{equation}
and
\begin{equation}
U_1(x_S, x_L) = U(x_S, x_L + h) \, .
\label{eq:pot_unbound_1}
\end{equation}
To connect the bound and unbound states, we consider the hybrid alchemical potential of Eq.\ (\ref{eq:pert_pot})
with the perturbation energy 
\begin{equation}
u_1(x_S, x_L) = U(x_S, x_L + h) - U(x_S, x_L) ,
\label{eq:pot_pert_change_1}
\end{equation}
which is defined as the change in the potential energy of the system for rigidly translating the ligand atoms from the binding site region to the bulk solvent region while all other degrees of freedom of the system remain unchanged.


With these definitions, the alchemical hybrid potential used for the first leg of the ATM cycle is:
\begin{equation}
{\rm Leg\ 1:} \quad U_\lambda(x_S, x_L) = U(x_S, x_L) + W_\lambda(u_1)  \, , \quad 0 <= \lambda <= 1/2
\label{eq:pot_alchemical_hybrid_sc_1}
\end{equation}
where the alchemical perturbation function $W_\lambda(u)$ is defined in Eq.\ (\ref{eq:ilog-function}) and the perturbation energy $u_1$ is defined in Eq.\ (\ref{eq:pot_pert_change_1}). As indicated in Eq.\ (\ref{eq:pot_alchemical_hybrid_sc_1}), the alchemical pathway for the first leg is terminated at $\lambda = 1/2$, where the ligand interacts with half strength with both the receptor environment and the solvent bulk. This ensures that severe steric clashes are not likely to occur at $\lambda = 1/2$. The $u_0$ parameter of the soft-core perturbation potential is set to a large enough value so that the perturbation potential $u$ does not exceed $u_0$ at $\lambda = 1/2$. Under these conditions, the original and the soft-core perturbation potentials are the same and it follows from Eq.\ (\ref{eq:pot_alchemical_hybrid_sc_1}) that the alchemical potential energy at $\lambda = 1/2$ is
\begin{equation}
U_{1/2}(x_S, x_L) = \frac12 U(x_S, x_L) + \frac12 U(x_S, x_L + h) .
\label{eq:pot_alchemical_hybrid_intermediate}
\end{equation}

Eq.\ (\ref{eq:pot_alchemical_hybrid_intermediate}) defines the potential energy function of the alchemical intermediate of the ATM cycle in Fig.\ \ref{fig:atm_vs_ddm}B. Accordingly, in the alchemical intermediate ensemble the ligand interacts symmetrically with the receptor and bulk solvent environments. Conversely, in this state the receptor atoms and the solvent molecules interact with the ligand at half strength. The alchemical calculation that corresponds to the alchemical potential Eq.\ (\ref{eq:pot_alchemical_hybrid_sc_1}) yields the free energy change $\Delta G_1$ in going from the bound state to the ATM alchemical intermediate.

To formulate the second leg of the ATM cycle connecting the unbound state to the intermediate state (Figure \ref{fig:atm_vs_ddm}), the role of the end states is reversed relative to the first leg. As before, $x_L$ describes the coordinates of the ligand in the receptor binding site and $x_L + h$ describes the coordinates of the ligand in the solvent bulk and the initial and final states are now defined by the potential energy functions of, respectively, the unbound and bound states:
\begin{equation}
U_0(x_S, x_L) = U(x_S, x_L + h)
\label{eq:pot_unbound_2}
\end{equation}
and
\begin{equation}
U_1(x_S, x_L) = U(x_S, x_L) \, .
\label{eq:pot_bound_2}
\end{equation}
The alchemical potential energy function for the second leg of the ATM cycle is
\begin{equation}
{\rm Leg\ 2:} \quad U_\lambda(x_S, x_L) = U(x_S, x_L + h) + W_\lambda(u_2) \, , \quad 0 <= \lambda <= 1/2
\label{eq:pot_alchemical_hybrid_sc_2}
\end{equation}
with the perturbation energy
\begin{equation}
    u_2(x_S, x_L) = U(x_S, x_L) - U(x_S, x_L + h)
\label{eq:pot_pert_change_2}
\end{equation}
that corresponds to the potential energy change of transferring the ligand from the bulk to the receptor binding site.
Note that under the same assumptions that led to Eq.\ (\ref{eq:pot_alchemical_hybrid_intermediate}),
Eq.\ (\ref{eq:pot_alchemical_hybrid_sc_2}) reaches $\lambda = 1/2$ at the same alchemical intermediate as Eq.\ (\ref{eq:pot_alchemical_hybrid_sc_1}). Thus, Eq.\ (\ref{eq:cycle_dg}) holds for the ATM thermodynamic cycle in Fig.\ \ref{fig:atm_vs_ddm}B. 

We end the presentation of the Alchemical Transfer Method by discussing the requirement of  splitting the alchemical path into two legs. Although the symmetric hybrid potential Eq.\ (\ref{eq:pot_alchemical_hybrid_intermediate}) can formally cover the direct path from the unbound to the bound states by extending the $\lambda$ range from $0$ to $1$, in practice it suffers severe end-point singularities at both of the end states. The potential energy, $U(x_S, x_L+h)$, when the ligand is placed in the solvent bulk region, is ill-defined near $\lambda = 0$ due to the atomic clashes that occur when ligand-solvent interactions are turned off. Conversely, near $\lambda = 1$, the potential energy $U(x_S, x_L)$, when the ligand is placed in the receptor site, is ill-defined due to clashes between ligand atoms and receptor atoms. The soft-core protocol we employ is based on an asymmetric definition of the perturbation potential [see Eqs.\ (\ref{eq:pot_pert_change_1}) and (\ref{eq:pot_pert_change_2})] that can address the singularity at one end-point or the other but not both simultaneously with only one continuous alchemical perturbation potential energy function.

\subsection{Software Implementation}

The method is implemented as an integrator plugin ({\tt github.com\-/rajatkrpal\-/openmm\_sdm\_plugin})\cite{pal2019perturbation} of the OpenMM library.\cite{eastman2017openmm} The integrator is based on the Langevin thermostat and high-level routines that displace the ligand, issue calls to energy and forces calculation routines, and compute the alchemical potential energy [Eq.\ (\ref{eq:ilog-function})] and its gradients by combining the returned system energies and forces. For example for leg 1, at each MD-step the plugin first computes and saves in temporary buffers the potential energy and the forces when the ligand is in the binding site. Then it displaces the ligand in the bulk by the displacement vector $h$ and recalculates the energy and forces (except for the binding site restraining potential, see below). The perturbation energy $u_1$ [Eq.\ (\ref{eq:pot_pert_change_1})] and its gradients are obtained by taking the corresponding differences before and after the displacement. The gradients of the alchemical perturbation energy (\ref{eq:pot_alchemical_hybrid_sc_1}) are derived from those of $u_1$ and of the undisplaced potential $U(x_S,x_L)$ by application of the chain rule. The resulting forces are then used to propagate atomic coordinates by one MD step. The same process is used for leg 2, except that the ligand is initially placed in the solvent bulk and it is translated into the binding site by reversing the displacement. In each case the binding restraint potential (see below) is applied when the ligand is in the binding site. No modifications of the core OpenMM energy routines are applied.

\subsection{Benchmark Systems}

The benchmark systems here were drawn from the host-guest systems presented in the SAMPL6 SAMPLing challenge\cite{rizzi2020sampl6}. The octa-acid (OA) and cucurbit[8]uril (CB8) hosts are well-studied supramolecular systems that have been featured in previous host-guest binding SAMPL challenges and the three guests resemble conventional druglike small molecules and fragments. In total, the benchmark systems presented here include 5-hexenoic acid (OA-G3) and 4-methylpentanoic acid (OA-G6) for the OA host, and quinine (CB3-G3) for the CB3 host. Despite the name, the guest G3 in CB8-G3 is distinct from the G3 guest in OA-G3. The parametrized systems, including their solvent descriptions, partial charges, and initial geometries, are provided at the github SAMPL6 site: {\tt github.com/\-samplchallenges/\-SAMPL6/\-tree/\-master/\-host\_guest/\-SAMPLing}.

Host-guest systems are practical alternatives to protein-ligand systems because of their minimal atom count and the improbability of undergoing major conformational reorganization. These systems thus enable the investigation of novel simulation techniques, such as longer timescales and appropriate sampling of multiple equivalent binding modes. Each host-guest complex has five conformations that differ in the position within the binding site, as well as in torsion angles. The procedure of obtaining five replicate free energy calculations allows for assessing the statistical uncertainty and reproducibility of the methodologies. The benchmark systems employed here create an accessible platform to improve both the predictive accuracy and computational efficiency of free energy calculations.\cite{rizzi2020sampl6}

\subsection{Computational Details}

The ATM calculations employed the host-guest  molecular complexes parametrized using the GAFF1.8/AM1-BCC force-field and the TIP3P water model as provided by Rizzi et al.\cite{rizzi2020sampl6} from {\tt(https://github.com\-/samplchallenges\-/SAMPL6\-/tree\-/master\-/host\_guest\-/SAMPLing)}. 
The Cartesian components of the displacement vector $h$ were set to approximately half the dimensions of the simulation box in order to ensure that the ligand is placed in the corner of the solvent box (Figure \ref{fig:cb8-g3}). This position is the farthest from the host, which the solvent box is centered around, and its periodic images. The complexes were energy minimized and thermalized at 300 K. Then, using the ATM alchemical potential energy function for leg 1  [Eq.\ (\ref{eq:pot_alchemical_hybrid_sc_1})] and starting at the bound state at $\lambda=0$, the systems were annealed to the symmetric intermediate $\lambda = 1/2$ for $250$ ps. The purpose of this step is to obtain a suitable initial configuration without severe unfavorable repulsion interactions at either end of the alchemical paths in order to start the molecular dynamics replica exchange alchemical calculations for each leg (see below).  
To limit the fluctuations of the position of the bulk solvent region, which would impact convergence, the position and orientation of the receptor was loosely restrained with a flat-bottom harmonic potential of force constant 25.0 kcal/(mol \AA$^2$) and a tolerance of 1.5 \AA\ on all of the heavy atoms of the lower portion of the receptor (the first 40 atoms of the host as listed in the provided files). 

Polar hydrogen atoms with zero Lennard-Jones parameters were modified to $\sigma_{\rm LJ} = 0.1$ \AA\ and $\epsilon_{\rm LJ} = 10^{-4}$ kcal/mol to avoid large attractive interactions between opposite charges at small distances in nearly uncoupled states. The change in potential energy of the system in the unbound, bound, and symmetric intermediate states due to this modification of the Lennard-Jones parameters is below single floating point precision. Single Decoupling alchemical calculations were prepared using the SDM workflow ({\tt github.com/\-egallicc/openmm\_sdm\_workflow.git}). MD calculations employed the OpenMM\cite{eastman2017openmm} MD engine and the SDM integrator plugins ({\tt github.com/\-rajatkrpal/\-openmm\_sdm\_plugin.git}) using the OpenCL platform. The ASyncRE software,\cite{gallicchio2015asynchronous} customized for OpenMM and SDM ({\tt github.com/\-egallicc/\-async\_re-openmm.git}), was used for the Hamiltonian Replica Exchange in $\lambda$ space for each ATM leg. 
 
The linear alchemical perturbation potential, $W_\lambda(u) = \lambda u_{\rm sc}(u)$,  corresponding to Eq.\ (\ref{eq:ilog-function}) with $\lambda_1 = \lambda_2 = \lambda$, was used for  the octa-acid (OA) systems with $11$ $\lambda$-states uniformly distributed between $\lambda=0$ and $1/2$ for each of the two ATM legs. The ATM calculations for the more challenging CB8-G3 complex employed the softplus perturbation potential Eq.\  (\ref{eq:ilog-function}) with $24$ $\lambda$-states and the parameters listed in Tables \ref{tab:CB8-g3-leg1} and \ref{tab:CB8-g3-leg2} for each leg. The parameters of the softplus perturbation potential were optimized using the scheme described previously\cite{pal2019perturbation,khuttan2021single} which involved running trial calculations with the linear potential, fitting the analytical model of binding\cite{kilburg2018analytical} to each transformation, and adjusting the parameters of the softplus potential to the resulting $\lambda$-function\cite{pal2019perturbation} to obtain a smooth alchemical transition. 

The soft-core perturbation energy Eq.\ (\ref{eq:soft-core-general}) was used for all calculations with $u_{\rm max}=300 $ kcal/mol, $u_0=100 $ kcal/mol.  The ligand was sequestered within the binding site by means of a flat-bottom harmonic potential between the centers of mass of the host and the ligand with a force constant of $25$ kcal/mol {\AA}$^2$ applied for separation greater than $4.5$ {\AA}. Perturbation energy samples and trajectory frames were saved every 5 ps. Hamiltonian replica exchanges in $\lambda$-space were performed every 5 ps. The Langevin thermostat with a time constant of 2 ps was used to maintain the temperature at 300 K. Each replica was simulated for a minimum of 10 ns. Binding free energies and the corresponding uncertainties were computed from the perturbation energy samples using UWHAM\cite{Tan2012}, discarding the first 5 ns of trajectory, followed by the addition of the concentration-dependent term $\Delta G^\circ_{\rm site} = -k_B T \ln  C^\circ V_{\rm site} = 0.87$ kcal/mol that corresponds to $300$ K temperature and the volume $V_{\rm site}$ of a sphere of radius $4.5$ \AA. The replica exchange simulations were run on the XSEDE Comet GPU HPC cluster at the San Diego Supercomputing Center each using four GPUs per node.

\begin{table}
  \begin{centering}
  \caption{\label{tab:CB8-g3-leg1} Alchemical schedule of the softplus perturbation function for leg 1 for the CB8/G3 complex.}
\begin{tabular}{lccccc}
$\lambda$  &   $\lambda_1$ & $\lambda_2$ & $\alpha$$^a$ & $u_0$$^b$ & $w_0$$^b$  \tabularnewline \hline
0.000 & 0.000 & 0.000 & 0.250 & 230.0 & 0.000 \tabularnewline
0.022 & 0.000 & 0.050 & 0.250 & 230.0 & 0.000 \tabularnewline
0.043 & 0.000 & 0.100 & 0.250 & 220.0 & 0.000 \tabularnewline
0.065 & 0.000 & 0.150 & 0.250 & 215.0 & 0.000 \tabularnewline
0.087 & 0.000 & 0.200 & 0.250 & 210.0 & 0.000 \tabularnewline
0.109 & 0.000 & 0.250 & 0.250 & 205.0 & 0.000 \tabularnewline
0.130 & 0.000 & 0.300 & 0.250 & 200.0 & 0.000 \tabularnewline
0.152 & 0.000 & 0.350 & 0.250 & 195.0 & 0.000 \tabularnewline
0.174 & 0.000 & 0.400 & 0.250 & 188.0 & 0.000 \tabularnewline
0.196 & 0.000 & 0.400 & 0.250 & 178.0 & 0.000 \tabularnewline
0.217 & 0.000 & 0.400 & 0.200 & 170.0 & 0.000 \tabularnewline
0.239 & 0.000 & 0.400 & 0.150 & 160.0 & 0.000 \tabularnewline
0.261 & 0.000 & 0.400 & 0.150 & 152.0 & 0.000 \tabularnewline
0.283 & 0.000 & 0.425 & 0.140 & 145.0 & 0.000 \tabularnewline
0.304 & 0.000 & 0.450 & 0.130 & 140.0 & 0.000 \tabularnewline
0.326 & 0.000 & 0.475 & 0.140 & 135.0 & 0.000 \tabularnewline
0.348 & 0.000 & 0.500 & 0.150 & 128.0 & 0.000 \tabularnewline
0.369 & 0.000 & 0.500 & 0.200 & 120.0 & 0.000 \tabularnewline
0.391 & 0.000 & 0.500 & 0.200 & 112.0 & 0.000 \tabularnewline
0.413 & 0.100 & 0.500 & 0.200 & 110.0 & 0.000 \tabularnewline
0.435 & 0.200 & 0.500 & 0.150 & 110.0 & 0.000 \tabularnewline
0.457 & 0.300 & 0.500 & 0.150 & 110.0 & 0.000 \tabularnewline
0.478 & 0.400 & 0.500 & 0.100 & 110.0 & 0.000 \tabularnewline
0.500 & 0.500 & 0.500 & 0.100 & 110.0 & 0.000 \tabularnewline \hline
\end{tabular} 
\end{centering}
\begin{flushleft}\small
$^a$In (kcal/mol)$^{-1}$. $^b$In kcal/mol.
\end{flushleft}
\end{table}

\begin{table}
\begin{centering}
  \caption{\label{tab:CB8-g3-leg2} Alchemical schedule of the softplus perturbation function for leg 2 for the CB8/G3 complex.}
\begin{tabular}{lccccc}
  $\lambda$  &   $\lambda_1$ & $\lambda_2$ & $\alpha$$^a$ & $u_0$$^b$ & $w_0$$^b$  \tabularnewline \hline
0.000 & 0.000 & 0.000 & 0.250 & 175 & 0.000 \tabularnewline
0.022 & 0.000 & 0.084 & 0.250 & 172.5 & 0.000 \tabularnewline
0.043 & 0.000 & 0.167 & 0.250 & 170 & 0.000 \tabularnewline
0.065 & 0.000 & 0.233 & 0.225 & 165 & 0.000 \tabularnewline
0.087 & 0.000 & 0.333 & 0.200 & 160 & 0.000 \tabularnewline
0.109 & 0.000 & 0.417 & 0.175 & 155 & 0.000 \tabularnewline
0.130 & 0.000 & 0.500 & 0.150 & 150 & 0.000 \tabularnewline
0.152 & 0.000 & 0.550 & 0.125 & 145 & 0.000 \tabularnewline
0.174 & 0.000 & 0.600 & 0.100 & 140 & 0.000 \tabularnewline
0.196 & 0.000 & 0.650 & 0.095 & 135 & 0.000 \tabularnewline
0.217 & 0.000 & 0.700 & 0.090 & 130 & 0.000 \tabularnewline
0.239 & 0.000 & 0.700 & 0.085 & 125 & 0.000 \tabularnewline
0.261 & 0.000 & 0.700 & 0.080 & 120 & 0.000 \tabularnewline
0.283 & 0.000 & 0.700 & 0.075 & 112 & 0.000 \tabularnewline
0.304 & 0.050 & 0.700 & 0.070 & 110 & 0.000 \tabularnewline
0.326 & 0.100 & 0.700 & 0.0675 & 102.5 & 0.000 \tabularnewline
0.348 & 0.150 & 0.700 & 0.065 & 105 & 0.000 \tabularnewline
0.369 & 0.200 & 0.700 & 0.060 & 100 & 0.000 \tabularnewline
0.391 & 0.250 & 0.650 & 0.055 & 100 & 0.000 \tabularnewline
0.413 & 0.300 & 0.625 & 0.050 & 100 & 0.000 \tabularnewline
0.435 & 0.350 & 0.600 & 0.045 & 100 & 0.000 \tabularnewline
0.457 & 0.400 & 0.575 & 0.040 & 100 & 0.000 \tabularnewline
0.478 & 0.450 & 0.550 & 0.035 & 100 & 0.000 \tabularnewline
0.500 & 0.500 & 0.500 & 0.030 & 100 & 0.000 \tabularnewline \hline
\end{tabular} 
\end{centering}
\begin{flushleft}\small
$^a$In (kcal/mol)$^{-1}$. $^b$In kcal/mol.
\end{flushleft}

\end{table}

\section{Results}

The standard binding free energy estimates, $\Delta G_b^\circ$, for the host-guest systems obtained using the ATM method are listed in Table \ref{tab:free_energies} compared to the corresponding experimental measurements, $\Delta G^\circ_b$(exp), and the reference computational estimates obtained through the attach-pull-release (APR) methodology.\cite{rizzi2020sampl6} 
Table \ref{tab:free_energies} also lists the number of energy and forces evaluations per replicate (a proxy for the computational cost) for the reference APR calculations, $n_{\rm eval}$(ref), and for the ATM calculations reported here ($n_{\rm eval}$). The APR method was selected as representative for this comparison because APR, similar to ATM and unlike DDM, displaces the ligand in a position in the solvent bulk at a finite distance from the host.

The ATM binding free energy estimates reported in Table \ref{tab:free_energies} are obtained as the average of five replicates started from different initial conformations as reported in Tables \ref{tab:OA-G3}, \ref{tab:OA-G6}, and \ref{tab:CB8-G3}. These tables also report the calculated free energies, $\Delta G_1$ and $\Delta G_2$, of the two alchemical legs for each replicate. The binding free energy of each replicate is the difference between those of the two legs (the excess component) plus the standard state term $\Delta G^\circ_{\rm site}$, which in this case measures out to be approximately $0.87$ (see Computational Details). The statistical uncertainties of the averages of each term reported in Tables \ref{tab:OA-G3}, \ref{tab:OA-G6}, and \ref{tab:CB8-G3} are expressed as 95\% confidence interval of the mean based on the t-test distribution with four degrees of freedom.\cite{rizzi2020sampl6} The statistical fluctuations of the binding free energies among the five conformations for each system were consistently smaller than the those of each of the legs, suggesting some level of systematic error cancellation. 

The ATM results obtained for the OA-G3 and OA-G6 complexes are in good agreement (within $0.5$ kcal/mol) with the reference values and well within the range of estimates obtained with other methods.\cite{rizzi2020sampl6} The ATM binding free energy estimate for the more challenging CB-G3 complex deviates more substantial from the APR reference ($2$ kcal/mol less favorable) and from those of the other methods applied to this benchmark system.\cite{rizzi2020sampl6} The origin of this discrepancy is not obvious. However all ATM estimates appear to generally underestimate binding affinities relative to the other methods bringing them, perhaps coincidentally, closer to the experimental measurements. The deviations between the experimental standard binding free energies and the ATM estimates are  $0.71$ kcal/mol, $1.35$ kcal/mol, and $2.08$ kcal/mol for, respectively, the OA-G3, OA-G6, and CB8-G3 complexes, compared to  $1.12$ kcal/mol, $1.83$ kcal/mol, and $4.05$ kcal/mol with APR. 

The range of the spread between ATM replicates obtained here for the octaacid systems is generally larger than with APR and other methods\cite{rizzi2020sampl6} albeit at a generally higher computational cost. For CB8-G3, ATM yields a spread similar to APR and the other methods with significantly less computational cost ($480$ vs.\ $2,135$ million energy evaluations as compared to APR, Table \ref{tab:free_energies}). 

In overall, ATM yields standard binding free energy estimates within the general range displayed by the established methods tested on the SAMPL6 SAMPLing benchmark set at a similar computational expense.\cite{rizzi2020sampl6} These initial results confirm the validity of the ATM approach.

\begin{table}
\begin{centering}
  \caption{\label{tab:free_energies} Standard binding free energy estimates and corresponding computational effort for the three host-guest complexes with the Alchemical Transfer Method compared to experimental and reference computed values.}
\begin{tabular}{lccccc}
  Complex & $\Delta G^\circ_b$(exp)$^{a,b}$  & $\Delta G^\circ_b$(ref)$^{a,c,d}$ & $n_{\rm eval}$(ref)$^{c,e}$   & $\Delta G^\circ_b$$^{a,e}$ & $n_{\rm eval}$$^f$\\ \hline
  OA-G3   & $-5.18 \pm 0.02$  & $ -6.3 \pm 0.1$   & $458\times10^6$  & $-5.89 \pm 0.33$ & $220\times10^6$ \\
  OA-G6   & $-4.97 \pm 0.02$  & $ -6.8 \pm 0.1$   & $305\times10^6$  & $-6.32 \pm 0.21$ & $220\times10^6$ \\
  CB8-G3  & $-6.45 \pm 0.06$  & $-10.5 \pm 0.6$   & $2135\times10^6$ & $-8.53 \pm 0.64$ & $480\times10^6$ \\ \hline
\end{tabular}
\end{centering}
\begin{flushleft}\small
$^a$In kcal/mol. $^b$From references \citenum{murkli2019cucurbit} and \citenum{sullivan2019thermodynamics}. $^c$From reference \citenum{rizzi2020sampl6}. $^d$APR method. $^e$This work, from Tables \ref{tab:OA-G3},  \ref{tab:OA-G6}, and \ref{tab:CB8-G3}. $^f$This work.
\end{flushleft}
\end{table}

\begin{table}
\begin{centering}
\caption{\label{tab:OA-G3} Free energy estimates for the two legs of the Alchemical Transfer Method and corresponding  standard binding free energy estimates for the OA-G3 complex starting with each of the the five initial SAMPL6 SAMPLing conformations. }
\begin{tabular}{lcccc}
     Conformation  &    $\Delta G_1$$^a$    & $\Delta G_2$$^a$         & $\Delta G^\circ_{\rm site}$$^a$ & $\Delta G^\circ_b$$^a$ \\ \hline 
     OA-G3-0 & $57.00$ & $50.47$ & $0.87$ & $-5.66$ \\ 
     OA-G3-1 & $57.79$ & $50.97$ & $0.87$ & $-5.95$ \\ 
     OA-G3-2 & $57.74$ & $50.79$ & $0.87$ & $-6.08$ \\ 
     OA-G3-3 & $57.65$ & $51.21$ & $0.87$ & $-5.57$ \\ 
     OA-G3-4 & $57.25$ & $50.19$ & $0.87$ & $-6.18$ \\ 
     Average$^b$ & $57.49 \pm 0.42$ & $50.73 \pm 0.50$ & $0.87$ & $-5.89 \pm 0.33$ \\ \hline
\end{tabular}
\end{centering}
\begin{flushleft}\small
$^a$In kcal/mol. $^b$With t-test 95\% confidence intervals with 4 degrees of freedom based on the standard deviation of the mean.
\end{flushleft}
\end{table}

\begin{table}
\begin{centering}
\caption{\label{tab:OA-G6} Free energy estimates for the two legs of the Alchemical Transfer Method and corresponding  standard binding free energy estimates for the OA-G6 complex starting with each of the the five initial SAMPL6 SAMPLing conformations.}
\begin{tabular}{lcccc}
 Conformation  &    $\Delta G_1$$^a$    & $\Delta G_2$$^a$         & $\Delta G^\circ_{\rm site}$$^a$ & $\Delta G^\circ_b$$^a$ \\ \hline 
  OA-G6-0 & $57.80$ & $50.64$ & $0.87$ & $-6.29$ \\ 
  OA-G6-1 & $58.19$ & $50.78$ & $0.87$ & $-6.54$ \\ 
  OA-G6-2 & $58.27$ & $50.97$ & $0.87$ & $-6.43$ \\ 
  OA-G6-3 & $58.74$ & $51.70$ & $0.87$ & $-6.17$ \\ 
  OA-G6-4 & $58.27$ & $51.25$ & $0.87$ & $-6.14$ \\
  Average$^b$ & $58.25 \pm 0.41$ & $51.07 \pm 0.52$ & $0.87$ & $-6.32 \pm 0.21$ \\ \hline
\end{tabular}
\end{centering}
\begin{flushleft}\small
$^a$In kcal/mol. $^b$With t-test 95\% confidence intervals with 4 degrees of freedom based on the standard deviation of the mean.
\end{flushleft}
\end{table}

\begin{table}
\begin{centering}
\caption{\label{tab:CB8-G3}  Free energy estimates for the two legs of the Alchemical Transfer Method and corresponding  standard binding free energy estimates for the CB8-G3 complex starting with each of the the five initial SAMPL6 SAMPLing conformations.}
\begin{tabular}{lcccc}
  Conformation  &    $\Delta G_1$$^a$    & $\Delta G_2$$^a$         & $\Delta G^\circ_{\rm site}$$^a$ & $\Delta G^\circ_b$$^a$ \\ \hline 
  CB8-G3-0 & $70.50$ & $61.06$ & $0.87$ & $-8.57$ \\ 
  CB8-G3-1 & $70.65$ & $60.81$ & $0.87$ & $-8.97$ \\
  CB8-G3-2 & $71.27$ & $61.40$ & $0.87$ & $-9.00$ \\ 
  CB8-G3-3 & $71.74$ & $62.96$ & $0.87$ & $-7.91$ \\ 
  CB8-G3-4 & $69.09$ & $60.02$ & $0.87$ & $-8.20$ \\
  Average &  $70.65\pm 1.24$  &  $61.25\pm 1.34$   & $0.87$  &$-8.53 \pm 0.64$   \\ \hline
\end{tabular}
\end{centering}
\begin{flushleft}\small
$^a$In kcal/mol. $^b$With t-test 95\% confidence intervals with 4 degrees of freedom based on the standard deviation of the mean.
\end{flushleft}
\end{table}

\section{Discussion}

The Alchemical Transfer Method (ATM) presented here implements a perturbation potential based on rigidly displacing the coordinates of the ligand atoms from a region in the solvent bulk to the receptor binding site or viceversa. Like in smart-darting Monte Carlo\cite{andricioaei2001smart}, this is accomplished using a displacement vector that can be thought as connecting a unique pair of points of two conformational macrostates. At each MD time-step the ligand disappears from one place and appears in another in a way that is physically not achievable or "alchemical". The change in potential energy of the system due to the ligand's displacement is the perturbation energy of the $\lambda$-dependent alchemical potential energy function that is used for conformational sampling and free energy estimation. While not presented here, the method is applicable to the calculation of the relative free energy of binding between two ligands\cite{Mey2020Best} by swapping their positions in the bulk and in the receptor site. This work is ongoing and will be reported in a forthcoming publication.

Similarly to physical pathway methods,\cite{Woo2005,henriksen2015computational,cruz2020combining} ATM is relatively easy to implement in molecular simulation packages because it does not require modifications of system parameters nor customized single- and dual-topologies setups that characterize conventional alchemical binding free energy methods. The method is illustrated here using a plugin implementation on top of the core OpenMM library.\cite{eastman2017openmm} The method is agnostic of the underlying energy function. It has been validated here with explicit solvation and Particle Mesh Ewald (PME) long range electrostatics. It is conceivably applicable without approximations to any kind of many-body potential function, including polarizable,\cite{harger2017tinker,panel2018accurate} quantum-mechanical,\cite{beierlein2011simple,lodola2012increasing,hudson2019use} and artificial neural network\cite{smith2019approaching,rufa2020towards} potentials. 

Unlike alchemical approaches such as double-decoupling,\cite{Gilson:Given:Bush:McCammon:97} which requires two systems, and dual-system single box alchemical methods,\cite{gapsys2015calculation,ekimoto2018elimination,procacci2020virtual} which require dual topologies, ATM works with a single standard model of the receptor-ligand complex solvated in a solvent box as in conventional molecular dynamics applications. 
In addition, ATM does not require soft-core pair potentials nor the splitting of the alchemical transformations into separate electrostatic and steric/dispersion steps.\cite{pal2019perturbation,khuttan2021single}

Similar to single-box alchemical approaches\cite{gapsys2015calculation,ekimoto2018elimination,procacci2020virtual} ATM avoids alchemical transformations that place the ligand in vacuum,\cite{Gilson:Given:Bush:McCammon:97,mobley2017predicting} which are particularly problematic for large and charged ligands.\cite{cruz2020combining}  The desolvation step of double-decoupling, for example, removes all of the ligand-water interactions, even though those of the solvent-exposed region of the ligand are likely to form again in the subsequent coupling step with the receptor. With ATM, instead, existing hydration interactions in the bulk are more likely to be replaced by similar interactions with the ligand displaced into the binding site. Moreover, unless the ligand is properly restrained, the vacuum intermediate is likely to introduce hard to converge free energy terms related to the reorganization of the ligand conformational ensemble from vacuum to the solvated environment.

ATM has some drawbacks, some of which are technical in nature and are likely to be addressed in the future. Because it calculates the system energy and forces twice for each MD time-step,\cite{Gallicchio2010} once with the ligand in the bulk and again with the ligand in the receptor pocket, the method is a factor of two slower per step than standard molecular dynamics. The two energy evaluations are however independent and can be conceivably run in parallel on two attached computational devices for added performance. Currently the method also requires the recalculation of the non-bonded neighbor list after each ligand displacement resulting in an additional 10 to 15\% slow-down per step with OpenMM for these systems. As we observed here for the CB8-G3 system, the binding of bulky ligands requires optimized softplus alchemical perturbation functions trained on trial calculations with the linear alchemical potential.\cite{kilburg2018analytical,pal2019perturbation,khuttan2021single} In future work, we plan to implement an adaptive algorithm to refine the parameters of the alchemical potential function on the fly.

Here we have validated ATM on the rigorous SAMPL6 SAMPLing dataset.\cite{rizzi2020sampl6} The dataset includes well-studied systems prepared with a single set of force field parameters and in different initial conformations to probe both systematic and statistical errors. The binding free energies of the systems have been computed and validated with a diverse collection of approaches, including alchemical and physical pathway methods.\cite{rizzi2020sampl6} The ATM results for the octacid systems obtained here are well within the range of estimates reported in reference \citenum{rizzi2020sampl6}, thereby adding confidence that the ATM approach is sound and that it has been implemented correctly. We observed in particular good agreement with the Attach Pull Release (APR) method,\cite{velez2013overcoming} a physical pathway approach in which the guest is progressively displaced into the solvent bulk to a comparable distance from the host as in this work. ATM yields a statistically significant less favorable binding free energy estimate than the other methods for the more challenging CB8-G3 system. The source of the deviation is unclear, however, ATM appears to generally yield binding free energies of smaller magnitude and closer to the experimental measurements than the other methods. Taking into account the relative computational expense, the statistical uncertainties obtained here indicate that ATM estimates have a comparable level of reproducibility and computational efficiency as the methods tested in reference \citenum{rizzi2020sampl6}.

\section{Conclusions}

We have presented the Alchemical Transfer Method (ATM) for the calculation of standard binding free energies of non-covalent molecular complexes. The method is based on a coordinate displacement perturbation of the ligand between the receptor binding site and the bulk solvent  and a thermodynamic cycle connected by a symmetric intermediate in which the ligand interacts with the receptor and solvent environments equally. While the approach is alchemical, ATM's implementation is as straightforward as physical pathway methods of binding. ATM does not require splitting the alchemical transformations into electrostatic and non-electrostatic steps and it does not employ soft-core pair potentials. We have implemented ATM as a freely available and open-source plugin of the OpenMM molecular dynamics library. The method and its implementation have been validated on the SAMPL6 SAMPLing host-guest benchmark set. 

\section{Acknowledgments}

 We acknowledge support from the National Science Foundation (NSF
CAREER 1750511 to E.G.). Molecular simulations were conducted on the
Comet GPU supercomputer cluster at the San Diego Supercomputing Center supported by NSF XSEDE award TG-MCB150001.

\providecommand{\latin}[1]{#1}
\makeatletter
\providecommand{\doi}
  {\begingroup\let\do\@makeother\dospecials
  \catcode`\{=1 \catcode`\}=2 \doi@aux}
\providecommand{\doi@aux}[1]{\endgroup\texttt{#1}}
\makeatother
\providecommand*\mcitethebibliography{\thebibliography}
\csname @ifundefined\endcsname{endmcitethebibliography}
  {\let\endmcitethebibliography\endthebibliography}{}


\providecommand{\latin}[1]{#1}
\makeatletter
\providecommand{\doi}
  {\begingroup\let\do\@makeother\dospecials
  \catcode`\{=1 \catcode`\}=2 \doi@aux}
\providecommand{\doi@aux}[1]{\endgroup\texttt{#1}}
\makeatother
\providecommand*\mcitethebibliography{\thebibliography}
\csname @ifundefined\endcsname{endmcitethebibliography}
  {\let\endmcitethebibliography\endthebibliography}{}
\begin{mcitethebibliography}{45}
\providecommand*\natexlab[1]{#1}
\providecommand*\mciteSetBstSublistMode[1]{}
\providecommand*\mciteSetBstMaxWidthForm[2]{}
\providecommand*\mciteBstWouldAddEndPuncttrue
  {\def\EndOfBibitem{\unskip.}}
\providecommand*\mciteBstWouldAddEndPunctfalse
  {\let\EndOfBibitem\relax}
\providecommand*\mciteSetBstMidEndSepPunct[3]{}
\providecommand*\mciteSetBstSublistLabelBeginEnd[3]{}
\providecommand*\EndOfBibitem{}
\mciteSetBstSublistMode{f}
\mciteSetBstMaxWidthForm{subitem}{(\alph{mcitesubitemcount})}
\mciteSetBstSublistLabelBeginEnd
  {\mcitemaxwidthsubitemform\space}
  {\relax}
  {\relax}

\bibitem[Simonson(2016)]{simonson2016physical}
Simonson,~T. The physical basis of ligand binding. \emph{In Silico Drug
  Discovery and Design} \textbf{2016}, 3--43\relax
\mciteBstWouldAddEndPuncttrue
\mciteSetBstMidEndSepPunct{\mcitedefaultmidpunct}
{\mcitedefaultendpunct}{\mcitedefaultseppunct}\relax
\EndOfBibitem
\bibitem[Jorgensen(2009)]{Jorgensen2009}
Jorgensen,~W.~L. Efficient drug lead discovery and optimization. \emph{Acc Chem
  Res} \textbf{2009}, \emph{42}, 724--733\relax
\mciteBstWouldAddEndPuncttrue
\mciteSetBstMidEndSepPunct{\mcitedefaultmidpunct}
{\mcitedefaultendpunct}{\mcitedefaultseppunct}\relax
\EndOfBibitem
\bibitem[Gallicchio and Levy(2011)Gallicchio, and Levy]{Gallicchio2011adv}
Gallicchio,~E.; Levy,~R.~M. Recent Theoretical and Computational Advances for
  Modeling Protein-Ligand Binding Affinities. \emph{Adv. Prot. Chem. Struct.
  Biol.} \textbf{2011}, \emph{85}, 27--80\relax
\mciteBstWouldAddEndPuncttrue
\mciteSetBstMidEndSepPunct{\mcitedefaultmidpunct}
{\mcitedefaultendpunct}{\mcitedefaultseppunct}\relax
\EndOfBibitem
\bibitem[Gallicchio \latin{et~al.}(2010)Gallicchio, Lapelosa, and
  Levy]{Gallicchio2010}
Gallicchio,~E.; Lapelosa,~M.; Levy,~R.~M. Binding Energy Distribution Analysis
  Method ({BEDAM}) for Estimation of Protein-Ligand Binding Affinities.
  \emph{J. Chem. Theory Comput.} \textbf{2010}, \emph{6}, 2961--2977\relax
\mciteBstWouldAddEndPuncttrue
\mciteSetBstMidEndSepPunct{\mcitedefaultmidpunct}
{\mcitedefaultendpunct}{\mcitedefaultseppunct}\relax
\EndOfBibitem
\bibitem[Pal \latin{et~al.}(2019)Pal, Ramsey, Gadhiya, Cordone, Wickstrom,
  Harding, Kurtzman, and Gallicchio]{pal2019dopamine}
Pal,~R.~K.; Ramsey,~S.; Gadhiya,~S.; Cordone,~P.; Wickstrom,~L.;
  Harding,~W.~W.; Kurtzman,~T.; Gallicchio,~E. Inclusion of Enclosed Hydration
  Effects in the Binding Free Energy Estimation of Dopamine {D3} Receptor
  Complexes. \emph{PLoS One} \textbf{2019}, \emph{14}, e0222092\relax
\mciteBstWouldAddEndPuncttrue
\mciteSetBstMidEndSepPunct{\mcitedefaultmidpunct}
{\mcitedefaultendpunct}{\mcitedefaultseppunct}\relax
\EndOfBibitem
\bibitem[Gilson \latin{et~al.}(1997)Gilson, Given, Bush, and
  McCammon]{Gilson:Given:Bush:McCammon:97}
Gilson,~M.~K.; Given,~J.~A.; Bush,~B.~L.; McCammon,~J.~A. The
  Statistical-Thermodynamic Basis for Computation of Binding Affinities: A
  Critical Review. \emph{Biophys. J.} \textbf{1997}, \emph{72},
  1047--1069\relax
\mciteBstWouldAddEndPuncttrue
\mciteSetBstMidEndSepPunct{\mcitedefaultmidpunct}
{\mcitedefaultendpunct}{\mcitedefaultseppunct}\relax
\EndOfBibitem
\bibitem[Deng \latin{et~al.}(2018)Deng, Cui, Zhang, Xia, Cruz, and
  Levy]{deng2018comparing}
Deng,~N.; Cui,~D.; Zhang,~B.~W.; Xia,~J.; Cruz,~J.; Levy,~R. Comparing
  alchemical and physical pathway methods for computing the absolute binding
  free energy of charged ligands. \emph{Phys. Chem. Chem. Phys.} \textbf{2018},
  \emph{20}, 17081--17092\relax
\mciteBstWouldAddEndPuncttrue
\mciteSetBstMidEndSepPunct{\mcitedefaultmidpunct}
{\mcitedefaultendpunct}{\mcitedefaultseppunct}\relax
\EndOfBibitem
\bibitem[Rocklin \latin{et~al.}(2013)Rocklin, Mobley, Dill, and
  H{\"u}nenberger]{rocklin2013calculating}
Rocklin,~G.~J.; Mobley,~D.~L.; Dill,~K.~A.; H{\"u}nenberger,~P.~H. Calculating
  the binding free energies of charged species based on explicit-solvent
  simulations employing lattice-sum methods: An accurate correction scheme for
  electrostatic finite-size effects. \emph{The Journal of chemical physics}
  \textbf{2013}, \emph{139}, 11B606\_1\relax
\mciteBstWouldAddEndPuncttrue
\mciteSetBstMidEndSepPunct{\mcitedefaultmidpunct}
{\mcitedefaultendpunct}{\mcitedefaultseppunct}\relax
\EndOfBibitem
\bibitem[Pan \latin{et~al.}(2017)Pan, Xu, Palpant, and
  Shaw]{pan2017quantitative}
Pan,~A.~C.; Xu,~H.; Palpant,~T.; Shaw,~D.~E. Quantitative characterization of
  the binding and unbinding of millimolar drug fragments with molecular
  dynamics simulations. \emph{J. Chem. Theory Comput.} \textbf{2017},
  \emph{13}, 3372--3377\relax
\mciteBstWouldAddEndPuncttrue
\mciteSetBstMidEndSepPunct{\mcitedefaultmidpunct}
{\mcitedefaultendpunct}{\mcitedefaultseppunct}\relax
\EndOfBibitem
\bibitem[\:{O}hlknecht \latin{et~al.}(2020)\:{O}hlknecht, Perthold, Lier, and
  Oostenbrink]{oostenbrink2020charge}
\:{O}hlknecht,~C.; Perthold,~J.~W.; Lier,~B.; Oostenbrink,~C. Charge-Changing
  Perturbations and Path Sampling via Classical Molecular Dynamic Simulations
  of Simple Guest–Host Systems. \emph{J. Chem. Theory Comput.} \textbf{2020},
  \emph{Article ASAP}\relax
\mciteBstWouldAddEndPuncttrue
\mciteSetBstMidEndSepPunct{\mcitedefaultmidpunct}
{\mcitedefaultendpunct}{\mcitedefaultseppunct}\relax
\EndOfBibitem
\bibitem[Lee \latin{et~al.}(2020)Lee, Lin, Allen, Lin, Radak, Tao, Tsai,
  Sherman, and York]{lee2020improved}
Lee,~T.-S.; Lin,~Z.; Allen,~B.~K.; Lin,~C.; Radak,~B.~K.; Tao,~Y.; Tsai,~H.-C.;
  Sherman,~W.; York,~D.~M. Improved Alchemical Free Energy Calculations with
  Optimized Smoothstep Softcore Potentials. \emph{J. Chem. Theory Comput.}
  \textbf{2020}, \relax
\mciteBstWouldAddEndPunctfalse
\mciteSetBstMidEndSepPunct{\mcitedefaultmidpunct}
{}{\mcitedefaultseppunct}\relax
\EndOfBibitem
\bibitem[Steinbrecher \latin{et~al.}(2007)Steinbrecher, Mobley, and
  Case]{Steinbrecher2007}
Steinbrecher,~T.; Mobley,~D.~L.; Case,~D.~A. Nonlinear scaling schemes for
  Lennard-Jones interactions in free energy calculations. \emph{J Chem Phys}
  \textbf{2007}, \emph{127}, 214108\relax
\mciteBstWouldAddEndPuncttrue
\mciteSetBstMidEndSepPunct{\mcitedefaultmidpunct}
{\mcitedefaultendpunct}{\mcitedefaultseppunct}\relax
\EndOfBibitem
\bibitem[Mey \latin{et~al.}(2020)Mey, Allen, Macdonald, Chodera, Hahn, Kuhn,
  Michel, Mobley, Naden, Prasad, Rizzi, Scheen, Shirts, Tresadern, and
  Xu]{Mey2020Best}
Mey,~A. S. J.~S.; Allen,~B.~K.; Macdonald,~H. E.~B.; Chodera,~J.~D.;
  Hahn,~D.~F.; Kuhn,~M.; Michel,~J.; Mobley,~D.~L.; Naden,~L.~N.; Prasad,~S.;
  Rizzi,~A.; Scheen,~J.; Shirts,~M.~R.; Tresadern,~G.; Xu,~H. Best Practices
  for Alchemical Free Energy Calculations [Article v1.0]. \emph{Living Journal
  of Computational Molecular Science} \textbf{2020}, \emph{2}, 18378\relax
\mciteBstWouldAddEndPuncttrue
\mciteSetBstMidEndSepPunct{\mcitedefaultmidpunct}
{\mcitedefaultendpunct}{\mcitedefaultseppunct}\relax
\EndOfBibitem
\bibitem[Lee \latin{et~al.}(2020)Lee, Allen, Giese, Guo, Li, Lin, McGee~Jr,
  Pearlman, Radak, Tao, Tsai, Xu, Sherman, and York]{lee2020alchemical}
Lee,~T.-S.; Allen,~B.~K.; Giese,~T.~J.; Guo,~Z.; Li,~P.; Lin,~C.;
  McGee~Jr,~T.~D.; Pearlman,~D.~A.; Radak,~B.~K.; Tao,~Y.; Tsai,~H.-C.; Xu,~H.;
  Sherman,~W.; York,~D.~M. Alchemical Binding Free Energy Calculations in
  {AMBER20}: Advances and Best Practices for Drug Discovery. \emph{J. Chem.
  Inf. Model.} \textbf{2020}, \relax
\mciteBstWouldAddEndPunctfalse
\mciteSetBstMidEndSepPunct{\mcitedefaultmidpunct}
{}{\mcitedefaultseppunct}\relax
\EndOfBibitem
\bibitem[Harger \latin{et~al.}(2017)Harger, Li, Wang, Dalby, Lagard{\`e}re,
  Piquemal, Ponder, and Ren]{harger2017tinker}
Harger,~M.; Li,~D.; Wang,~Z.; Dalby,~K.; Lagard{\`e}re,~L.; Piquemal,~J.-P.;
  Ponder,~J.; Ren,~P. Tinker-OpenMM: Absolute and relative alchemical free
  energies using AMOEBA on GPUs. \emph{J. Comp. Chem.} \textbf{2017},
  \emph{38}, 2047--2055\relax
\mciteBstWouldAddEndPuncttrue
\mciteSetBstMidEndSepPunct{\mcitedefaultmidpunct}
{\mcitedefaultendpunct}{\mcitedefaultseppunct}\relax
\EndOfBibitem
\bibitem[Darden \latin{et~al.}(1993)Darden, York, and Pedersen]{Darden:93}
Darden,~T.~A.; York,~D.~M.; Pedersen,~L.~G. \emph{J. Chem. Phys.}
  \textbf{1993}, \emph{98}, 10089--10092\relax
\mciteBstWouldAddEndPuncttrue
\mciteSetBstMidEndSepPunct{\mcitedefaultmidpunct}
{\mcitedefaultendpunct}{\mcitedefaultseppunct}\relax
\EndOfBibitem
\bibitem[Gallicchio \latin{et~al.}(2009)Gallicchio, Paris, and
  Levy]{Gallicchio2009}
Gallicchio,~E.; Paris,~K.; Levy,~R.~M. The {AGBNP2} Implicit Solvation Model.
  \emph{J. Chem. Theory Comput.} \textbf{2009}, \emph{5}, 2544--2564\relax
\mciteBstWouldAddEndPuncttrue
\mciteSetBstMidEndSepPunct{\mcitedefaultmidpunct}
{\mcitedefaultendpunct}{\mcitedefaultseppunct}\relax
\EndOfBibitem
\bibitem[Zou \latin{et~al.}(2019)Zou, Tian, and Simmerling]{zou2019blinded}
Zou,~J.; Tian,~C.; Simmerling,~C. Blinded prediction of protein--ligand binding
  affinity using Amber thermodynamic integration for the 2018 D3R grand
  challenge 4. \emph{J. Comp. Aid. Mol. Des.} \textbf{2019}, \emph{33},
  1021--1029\relax
\mciteBstWouldAddEndPuncttrue
\mciteSetBstMidEndSepPunct{\mcitedefaultmidpunct}
{\mcitedefaultendpunct}{\mcitedefaultseppunct}\relax
\EndOfBibitem
\bibitem[Woo and Roux(2005)Woo, and Roux]{Woo2005}
Woo,~H.-J.; Roux,~B. Calculation of absolute protein-ligand binding free energy
  from computer simulations. \emph{Proc. Natl. Acad. Sci. USA} \textbf{2005},
  \emph{102}, 6825--6830\relax
\mciteBstWouldAddEndPuncttrue
\mciteSetBstMidEndSepPunct{\mcitedefaultmidpunct}
{\mcitedefaultendpunct}{\mcitedefaultseppunct}\relax
\EndOfBibitem
\bibitem[Limongelli \latin{et~al.}(2013)Limongelli, Bonomi, and
  Parrinello]{limongelli2013funnel}
Limongelli,~V.; Bonomi,~M.; Parrinello,~M. Funnel metadynamics as accurate
  binding free-energy method. \emph{Proc. Natl. Acad. Sci.} \textbf{2013},
  \emph{110}, 6358--6363\relax
\mciteBstWouldAddEndPuncttrue
\mciteSetBstMidEndSepPunct{\mcitedefaultmidpunct}
{\mcitedefaultendpunct}{\mcitedefaultseppunct}\relax
\EndOfBibitem
\bibitem[Cruz \latin{et~al.}(2020)Cruz, Wickstrom, Yang, Gallicchio, and
  Deng]{cruz2020combining}
Cruz,~J.; Wickstrom,~L.; Yang,~D.; Gallicchio,~E.; Deng,~N. Combining
  Alchemical Transformation with a Physical Pathway to Accelerate Absolute
  Binding Free Energy Calculations of Charged Ligands to Enclosed Binding
  Sites. \emph{J. Chem. Theory Comput.} \textbf{2020}, \emph{16},
  2803--2813\relax
\mciteBstWouldAddEndPuncttrue
\mciteSetBstMidEndSepPunct{\mcitedefaultmidpunct}
{\mcitedefaultendpunct}{\mcitedefaultseppunct}\relax
\EndOfBibitem
\bibitem[Kilburg and Gallicchio(2018)Kilburg, and
  Gallicchio]{kilburg2018assessment}
Kilburg,~D.; Gallicchio,~E. Assessment of a Single Decoupling Alchemical
  Approach for the Calculation of the Absolute Binding Free Energies of
  Protein-Peptide Complexes. \emph{Frontiers in Molecular Biosciences}
  \textbf{2018}, \emph{5}, 22\relax
\mciteBstWouldAddEndPuncttrue
\mciteSetBstMidEndSepPunct{\mcitedefaultmidpunct}
{\mcitedefaultendpunct}{\mcitedefaultseppunct}\relax
\EndOfBibitem
\bibitem[Pal and Gallicchio(2019)Pal, and Gallicchio]{pal2019perturbation}
Pal,~R.~K.; Gallicchio,~E. Perturbation potentials to overcome order/disorder
  transitions in alchemical binding free energy calculations. \emph{J. Chem.
  Phys.} \textbf{2019}, \emph{151}, 124116\relax
\mciteBstWouldAddEndPuncttrue
\mciteSetBstMidEndSepPunct{\mcitedefaultmidpunct}
{\mcitedefaultendpunct}{\mcitedefaultseppunct}\relax
\EndOfBibitem
\bibitem[Eastman \latin{et~al.}(2017)Eastman, Swails, Chodera, McGibbon, Zhao,
  Beauchamp, Wang, Simmonett, Harrigan, Stern, \latin{et~al.}
  others]{eastman2017openmm}
others,, \latin{et~al.}  OpenMM 7: Rapid development of high performance
  algorithms for molecular dynamics. \emph{PLoS Comp. Bio.} \textbf{2017},
  \emph{13}, e1005659\relax
\mciteBstWouldAddEndPuncttrue
\mciteSetBstMidEndSepPunct{\mcitedefaultmidpunct}
{\mcitedefaultendpunct}{\mcitedefaultseppunct}\relax
\EndOfBibitem
\bibitem[Khuttan \latin{et~al.}(2021)Khuttan, Azimi, Wu, and
  Gallicchio]{khuttan2021single}
Khuttan,~S.; Azimi,~S.; Wu,~J.~Z.; Gallicchio,~E. Alchemical Transformations
  for Concerted Hydration Free Energy Estimation with Explicit Solvation.
  \emph{J. Chem. Phys} \textbf{2021}, In press.\relax
\mciteBstWouldAddEndPunctfalse
\mciteSetBstMidEndSepPunct{\mcitedefaultmidpunct}
{}{\mcitedefaultseppunct}\relax
\EndOfBibitem
\bibitem[Rizzi \latin{et~al.}(2020)Rizzi, Jensen, Slochower, Aldeghi, Gapsys,
  Ntekoumes, Bosisio, Papadourakis, Henriksen, De~Groot, Cournia, Dickson,
  Michel, Gilson, Shirts, Mobley, and Chodera]{rizzi2020sampl6}
Rizzi,~A. \latin{et~al.}  The {SAMPL6 SAMPLing} challenge: Assessing the
  reliability and efficiency of binding free energy calculations. \emph{J.
  Comp. Aid. Mol. Des.} \textbf{2020}, 1--33\relax
\mciteBstWouldAddEndPuncttrue
\mciteSetBstMidEndSepPunct{\mcitedefaultmidpunct}
{\mcitedefaultendpunct}{\mcitedefaultseppunct}\relax
\EndOfBibitem
\bibitem[Gallicchio \latin{et~al.}(2015)Gallicchio, Xia, Flynn, Zhang,
  Samlalsingh, Mentes, and Levy]{gallicchio2015asynchronous}
Gallicchio,~E.; Xia,~J.; Flynn,~W.~F.; Zhang,~B.; Samlalsingh,~S.; Mentes,~A.;
  Levy,~R.~M. Asynchronous replica exchange software for grid and heterogeneous
  computing. \emph{Computer Physics Communications} \textbf{2015}, \emph{196},
  236--246\relax
\mciteBstWouldAddEndPuncttrue
\mciteSetBstMidEndSepPunct{\mcitedefaultmidpunct}
{\mcitedefaultendpunct}{\mcitedefaultseppunct}\relax
\EndOfBibitem
\bibitem[Kilburg and Gallicchio(2018)Kilburg, and
  Gallicchio]{kilburg2018analytical}
Kilburg,~D.; Gallicchio,~E. Analytical Model of the Free Energy of Alchemical
  Molecular Binding. \emph{J. Chem. Theory Comput.} \textbf{2018}, \emph{14},
  6183--6196\relax
\mciteBstWouldAddEndPuncttrue
\mciteSetBstMidEndSepPunct{\mcitedefaultmidpunct}
{\mcitedefaultendpunct}{\mcitedefaultseppunct}\relax
\EndOfBibitem
\bibitem[Tan \latin{et~al.}(2012)Tan, Gallicchio, Lapelosa, and Levy]{Tan2012}
Tan,~Z.; Gallicchio,~E.; Lapelosa,~M.; Levy,~R.~M. Theory of binless
  multi-state free energy estimation with applications to protein-ligand
  binding. \emph{J. Chem. Phys.} \textbf{2012}, \emph{136}, 144102\relax
\mciteBstWouldAddEndPuncttrue
\mciteSetBstMidEndSepPunct{\mcitedefaultmidpunct}
{\mcitedefaultendpunct}{\mcitedefaultseppunct}\relax
\EndOfBibitem
\bibitem[Murkli \latin{et~al.}(2019)Murkli, McNeill, and
  Isaacs]{murkli2019cucurbit}
Murkli,~S.; McNeill,~J.~N.; Isaacs,~L. Cucurbit[8]uril guest complexes: blinded
  dataset for the SAMPL6 challenge. \emph{Supramolecular Chemistry}
  \textbf{2019}, \emph{31}, 150--158\relax
\mciteBstWouldAddEndPuncttrue
\mciteSetBstMidEndSepPunct{\mcitedefaultmidpunct}
{\mcitedefaultendpunct}{\mcitedefaultseppunct}\relax
\EndOfBibitem
\bibitem[Sullivan \latin{et~al.}(2019)Sullivan, Yao, and
  Gibb]{sullivan2019thermodynamics}
Sullivan,~M.~R.; Yao,~W.; Gibb,~B.~C. The thermodynamics of guest complexation
  to octa-acid and tetra-endo-methyl octa-acid: reference data for the sixth
  statistical assessment of modeling of proteins and ligands {(SAMPL6)}.
  \emph{Supramolecular chemistry} \textbf{2019}, \emph{31}, 184--189\relax
\mciteBstWouldAddEndPuncttrue
\mciteSetBstMidEndSepPunct{\mcitedefaultmidpunct}
{\mcitedefaultendpunct}{\mcitedefaultseppunct}\relax
\EndOfBibitem
\bibitem[Andricioaei \latin{et~al.}(2001)Andricioaei, Straub, and
  Voter]{andricioaei2001smart}
Andricioaei,~I.; Straub,~J.~E.; Voter,~A.~F. Smart darting monte carlo.
  \emph{J. Chem. Phys.} \textbf{2001}, \emph{114}, 6994--7000\relax
\mciteBstWouldAddEndPuncttrue
\mciteSetBstMidEndSepPunct{\mcitedefaultmidpunct}
{\mcitedefaultendpunct}{\mcitedefaultseppunct}\relax
\EndOfBibitem
\bibitem[Henriksen \latin{et~al.}(2015)Henriksen, Fenley, and
  Gilson]{henriksen2015computational}
Henriksen,~N.~M.; Fenley,~A.~T.; Gilson,~M.~K. Computational calorimetry:
  high-precision calculation of host--guest binding thermodynamics. \emph{J.
  Chem. Theory Comput.} \textbf{2015}, \emph{11}, 4377--4394\relax
\mciteBstWouldAddEndPuncttrue
\mciteSetBstMidEndSepPunct{\mcitedefaultmidpunct}
{\mcitedefaultendpunct}{\mcitedefaultseppunct}\relax
\EndOfBibitem
\bibitem[Panel \latin{et~al.}(2018)Panel, Villa, Fuentes, and
  Simonson]{panel2018accurate}
Panel,~N.; Villa,~F.; Fuentes,~E.~J.; Simonson,~T. Accurate PDZ/peptide binding
  specificity with additive and polarizable free energy simulations.
  \emph{Biophys. J.} \textbf{2018}, \emph{114}, 1091--1102\relax
\mciteBstWouldAddEndPuncttrue
\mciteSetBstMidEndSepPunct{\mcitedefaultmidpunct}
{\mcitedefaultendpunct}{\mcitedefaultseppunct}\relax
\EndOfBibitem
\bibitem[Beierlein \latin{et~al.}(2011)Beierlein, Michel, and
  Essex]{beierlein2011simple}
Beierlein,~F.~R.; Michel,~J.; Essex,~J.~W. A simple {QM/MM} approach for
  capturing polarization effects in protein- ligand binding free energy
  calculations. \emph{J. Phys. Chem. B} \textbf{2011}, \emph{115},
  4911--4926\relax
\mciteBstWouldAddEndPuncttrue
\mciteSetBstMidEndSepPunct{\mcitedefaultmidpunct}
{\mcitedefaultendpunct}{\mcitedefaultseppunct}\relax
\EndOfBibitem
\bibitem[Lodola and De~Vivo(2012)Lodola, and De~Vivo]{lodola2012increasing}
Lodola,~A.; De~Vivo,~M. \emph{Adv. Prot. Chem. Struct. Biol.}; Elsevier, 2012;
  Vol.~87; pp 337--362\relax
\mciteBstWouldAddEndPuncttrue
\mciteSetBstMidEndSepPunct{\mcitedefaultmidpunct}
{\mcitedefaultendpunct}{\mcitedefaultseppunct}\relax
\EndOfBibitem
\bibitem[Hudson \latin{et~al.}(2019)Hudson, Woodcock, and
  Boresch]{hudson2019use}
Hudson,~P.~S.; Woodcock,~H.~L.; Boresch,~S. Use of interaction energies in
  QM/MM free energy simulations. \emph{J. Chem. Theory Comput.} \textbf{2019},
  \emph{15}, 4632--4645\relax
\mciteBstWouldAddEndPuncttrue
\mciteSetBstMidEndSepPunct{\mcitedefaultmidpunct}
{\mcitedefaultendpunct}{\mcitedefaultseppunct}\relax
\EndOfBibitem
\bibitem[Smith \latin{et~al.}(2019)Smith, Nebgen, Zubatyuk, Lubbers, Devereux,
  Barros, Tretiak, Isayev, and Roitberg]{smith2019approaching}
Smith,~J.~S.; Nebgen,~B.~T.; Zubatyuk,~R.; Lubbers,~N.; Devereux,~C.;
  Barros,~K.; Tretiak,~S.; Isayev,~O.; Roitberg,~A.~E. Approaching coupled
  cluster accuracy with a general-purpose neural network potential through
  transfer learning. \emph{Nature Comm.} \textbf{2019}, \emph{10}, 1--8\relax
\mciteBstWouldAddEndPuncttrue
\mciteSetBstMidEndSepPunct{\mcitedefaultmidpunct}
{\mcitedefaultendpunct}{\mcitedefaultseppunct}\relax
\EndOfBibitem
\bibitem[Rufa \latin{et~al.}(2020)Rufa, Macdonald, Fass, Wieder, Grinaway,
  Roitberg, Isayev, and Chodera]{rufa2020towards}
Rufa,~D.~A.; Macdonald,~H. E.~B.; Fass,~J.; Wieder,~M.; Grinaway,~P.~B.;
  Roitberg,~A.~E.; Isayev,~O.; Chodera,~J.~D. Towards chemical accuracy for
  alchemical free energy calculations with hybrid physics-based machine
  learning/molecular mechanics potentials. \emph{BioRxiv} \textbf{2020}, \relax
\mciteBstWouldAddEndPunctfalse
\mciteSetBstMidEndSepPunct{\mcitedefaultmidpunct}
{}{\mcitedefaultseppunct}\relax
\EndOfBibitem
\bibitem[Gapsys \latin{et~al.}(2015)Gapsys, Michielssens, Peters, de~Groot, and
  Leonov]{gapsys2015calculation}
Gapsys,~V.; Michielssens,~S.; Peters,~J.~H.; de~Groot,~B.~L.; Leonov,~H.
  \emph{Molecular Modeling of Proteins}; Springer, 2015; pp 173--209\relax
\mciteBstWouldAddEndPuncttrue
\mciteSetBstMidEndSepPunct{\mcitedefaultmidpunct}
{\mcitedefaultendpunct}{\mcitedefaultseppunct}\relax
\EndOfBibitem
\bibitem[Ekimoto \latin{et~al.}(2018)Ekimoto, Yamane, and
  Ikeguchi]{ekimoto2018elimination}
Ekimoto,~T.; Yamane,~T.; Ikeguchi,~M. Elimination of finite-size effects on
  binding free energies via the warp-drive method. \emph{Journal of chemical
  theory and computation} \textbf{2018}, \emph{14}, 6544--6559\relax
\mciteBstWouldAddEndPuncttrue
\mciteSetBstMidEndSepPunct{\mcitedefaultmidpunct}
{\mcitedefaultendpunct}{\mcitedefaultseppunct}\relax
\EndOfBibitem
\bibitem[Macchiagodena \latin{et~al.}(2020)Macchiagodena, Pagliai, Karrenbrock,
  Guarnieri, Iannone, and Procacci]{procacci2020virtual}
Macchiagodena,~M.; Pagliai,~M.; Karrenbrock,~M.; Guarnieri,~G.; Iannone,~F.;
  Procacci,~P. Virtual Double-System Single-Box: A Nonequilibrium Alchemical
  Technique for Absolute Binding Free Energy Calculations: Application to
  Ligands of the {SARS-CoV-2 Main Protease}. \emph{J. Chem. Theory Comput.}
  \textbf{2020}, \relax
\mciteBstWouldAddEndPunctfalse
\mciteSetBstMidEndSepPunct{\mcitedefaultmidpunct}
{}{\mcitedefaultseppunct}\relax
\EndOfBibitem
\bibitem[Mobley and Gilson(2017)Mobley, and Gilson]{mobley2017predicting}
Mobley,~D.~L.; Gilson,~M.~K. Predicting binding free energies: frontiers and
  benchmarks. \emph{Ann. Rev. Bioph.} \textbf{2017}, \emph{46}, 531--558\relax
\mciteBstWouldAddEndPuncttrue
\mciteSetBstMidEndSepPunct{\mcitedefaultmidpunct}
{\mcitedefaultendpunct}{\mcitedefaultseppunct}\relax
\EndOfBibitem
\bibitem[Velez-Vega and Gilson(2013)Velez-Vega, and
  Gilson]{velez2013overcoming}
Velez-Vega,~C.; Gilson,~M.~K. Overcoming dissipation in the calculation of
  standard binding free energies by ligand extraction. \emph{J. Comp. Chem.}
  \textbf{2013}, \emph{34}, 2360--2371\relax
\mciteBstWouldAddEndPuncttrue
\mciteSetBstMidEndSepPunct{\mcitedefaultmidpunct}
{\mcitedefaultendpunct}{\mcitedefaultseppunct}\relax
\EndOfBibitem
\end{mcitethebibliography}

\end{document}